\documentclass[aps,prl,reprint,superscriptaddress]{revtex4-1}

\usepackage{graphicx}
\usepackage{rotating}
\usepackage{amssymb}    
\usepackage{amsmath}   
\usepackage{epsfig}
\usepackage[normalem]{ulem}   
\usepackage{bm}   
\usepackage{color}
\usepackage{dcolumn}

\date{\today}

\begin{document}

\title{Geomagnetic reversals at the edge of regularity}

\author{Breno Raphaldini}
\affiliation{Instituto de Astronomia Geof\'isica e Ci\^encias Atmosf\'ericas - Universidade de S\~ao Paulo, 05508-090 S\~ao Paulo, Brazil}

\author{Everton S. Medeiros}
\affiliation{Institut f\"ur Theoretische Physik, Technische Universit\"at Berlin, Hardenbergstra\ss e 36, 10623 Berlin, Germany}
 
\author{David Ciro}
\affiliation{Centro de Ci{\^e}ncias Biol{\'o}gicas, Universidade Federal de Santa Catarina, Florian{\'o}polis, Brazil}

\author{Daniel Ribeiro Franco}
\affiliation{Observatorio Nacional, 04533-085, Rio de Janeiro, Brazil}

\author{Ricardo Ivan Ferreira Trindade}
\affiliation{Instituto de Astronomia Geof\'isica e Ci\^encias Atmosf\'ericas - Universidade de S\~ao Paulo, 05508-090 S\~ao Paulo, Brazil}

\begin{abstract}
 Geomagnetic field reversals remain as one of the most intriguing problems in geophysics and are regarded as chaotic processes resulting from a dynamo mechanism. In this article, we use the polarity scale data set for the last 170 Myr from the ocean's floor to provide robust evidence for an inverse relationship between the complexity of sequences of consecutive polarity intervals and the respective reversal rate. In particular the variability of sequences of polarity intervals reaches a minimum in the mid-Jurassic when a maximum reversal rate is found. This raises the possibility of epochs of high regularity in the reversal process geodynamo. To shed light on this process, We investigate the transition from regular to chaotic regime in a minimal model for geomagnetic reversals. We show that even in a chaotic regime, near to the transition the system retains the signature of regular behavior. We suggest that geomagnetic reversals have switched between different degrees of irregularity,  with a dominant periodicity of $\approx 70$ kyrs that results from a “ghost limit cycle”.
\end{abstract}
\maketitle

The geomagnetic field is characterized by a dominant dipole component that reverses its polarity in irregular times. The physical mechanism driving reversals of the geomagnetic field's dipole is still not well understood, although it's well established that reversals are linked to the dynamo process that takes place on Earth's outer core \cite{roberts2000geodynamo}. It is hypothesized that long term changes in the reversal processes must be linked to the evolution of the dynamo process in the Earth's outer core. For instance the growth of the inner core, \cite{tarduno2006paleomagnetism}, and most notably the changes in the properties of the core-mantle boundary could have large impacts in the reversal process, \cite{amit2015lower,bloxham2000sensitivity}.

Polarity scales compiled from paleomagnetic data reveal a large variability of the polarity intervals, \cite{cande1995revised, gradstein2012geologic}. Short intervals have a duration of the order of tens of thousands of years, while exceptionally long intervals with a single polarity last $O(10^7)$ yrs and are called Superchrons.

The statistics and variability of reversal rates is a matter of debate. It was long assumed the reversal rates to follow a Poisson type of renewal process. This leads to an exponential distribution, $p(T) \propto exp(-\lambda T)$, where $\lambda$ is the rate of the process. This type of process is memoryless leading to an independent sequence of polarity intervals, see \cite{cox1968lengths, mcelhinny1998magnetic}. Subsequent articles investigated the hypothesis of a Poisson process in which the rate itself evolves in time $\lambda=\lambda(t)$ \cite{constable2000rates}. \citet{carbone2006clustering}, later on, showed that the sequences of reversals largely departs from a Poisson process. They suggested that reversals are better modeled by a Levy-type process. This result contrasts from previous research, implying in long-range correlations between intervals duration, with different classes of intervals clustered in time. However, the physical mechanism behind these long-range dependencies remains elusive. Insights are provided from other natural dynamos, such as the Sun and other stars usually operating in a quasi-periodic cyclic manner. The Sun has a well-recorded $11$ year cycle, named Schwabe cycle, with peaks in sunspot numbers each $11$ years on average occurring in anti-phase with its axial dipole dynamics. See \cite{usoskin2017history} for a review. This analogy raise the question of whether the geodynamo operates in more regular regimes similar to other natural dynamos.

In this Letter, we address this issue by searching for regularities in the sequence of geomagnetic reversals recorded in the last $170$ Myrs. Specifically, we assess the level of regularity in this signal by estimating its sample entropy (SamEn) and coefficient of variation (C). With this, we provide statistical evidence for significant variations in the signal's regularity at the time scale of $10^7$ yrs. And, more importantly, we find a period of highly regular reversals around $160$ Myrs ago. The overall irregularity of the signal is evidenced in the large variability in the density polarity intervals, however, we observe preferential intervals pointing out to underlying periodic processes. Finally, we interpret these results in the framework of nonlinear dynamics by analyzing the transitions from regular to chaotic behavior in a simple model for geomagnetic reversals.

To quantify the degree of regularity in the geomagnetic reversals, we invoke the concept of chaos from nonlinear dynamics. Chaos is usually characterized by the exponential separation of nearby trajectories in the system's state-space as the time evolves \cite{ruelle1989chaotic}. The rate of such separation is quantified by the Lyapunov spectrum $\lambda_i(x)$ \cite{eckmann1985ergodic}. Another way of measuring the complexity of a dynamical system is the Kolmogorov-Sinai (KS) entropy, that qualitatively measures the rate of creation of information of the system \cite{eckmann1985ergodic,viana2016foundations}. For a class of dynamical systems \citet{ruelle1989chaotic}, the KS entropy and the Lyapunov spectrum are linked by the Pesin identity:
\begin{equation}
    \mathcal{S}_{KS}= \int \sum_{\lambda>0}\lambda_i(x) d\chi,
    \label{KS_entropy}
\end{equation}
where  $\chi$ is the ergodic invariant probability measure of the chaotic attractor. The KS entropy is therefore the sum of all positive Lyapunov exponents of the system. 
 
For applications to real-world data, the definition (\ref{KS_entropy}) needs to be adapted to overpass intrinsic difficulties arising from observational procedures. For this task, a measure called approximate entropy (ApEn) \cite{pincus1991approximate} and its successor sample entropy (SamEn) \cite{richman2000physiological} have been successfully employed to quantify the amount of regularity in electrocardiograms \cite{pincus1991regularity}, and cardiovascular signals in general  \cite{porta2018relevance}. Here, we implement SamEn to assess the level of regularities in the dynamics of reversals of the geomagnetic field. Given a data sample $X=(x_1,..., x_N)$, a subset of $X$ of size $m$ is written as $X_i=(x_i, ..., x_{i+m-1})$ and the corresponding shifted subset as $X_{i+1}=(x_{i+1}, ..., x_{i+m})$ with $1\leq i\leq i+m\leq N$. The SamEn is defined as the negative logarithm of the probability that the Euclidean distance is less than a constant, i.e., $d(X_i,X_{i+1})<r$. In our analysis, we adopt commonly used values of $r$ and $m$, namely, $m=2$ and $r=0.2 \sigma (X)$, where $\sigma (X)$ denotes the standard deviation of the sample $X$. We complement our analysis on the dispersion of the persistence times by calculating the coefficient of variation $C=\sigma (X)/E(X)$, where $E$ is the expected value. The coefficient of variation has an important implication for our analysis, since for any random variable $X$ with an exponential distribution $C(X)=1$ \cite{Note1}, this enables us to evaluate the Poissonian hypothesis \cite{mcelhinny1998magnetic,constable2000rates}.   

\begin{figure}[ht]
\centering\includegraphics[width=1.0\linewidth]{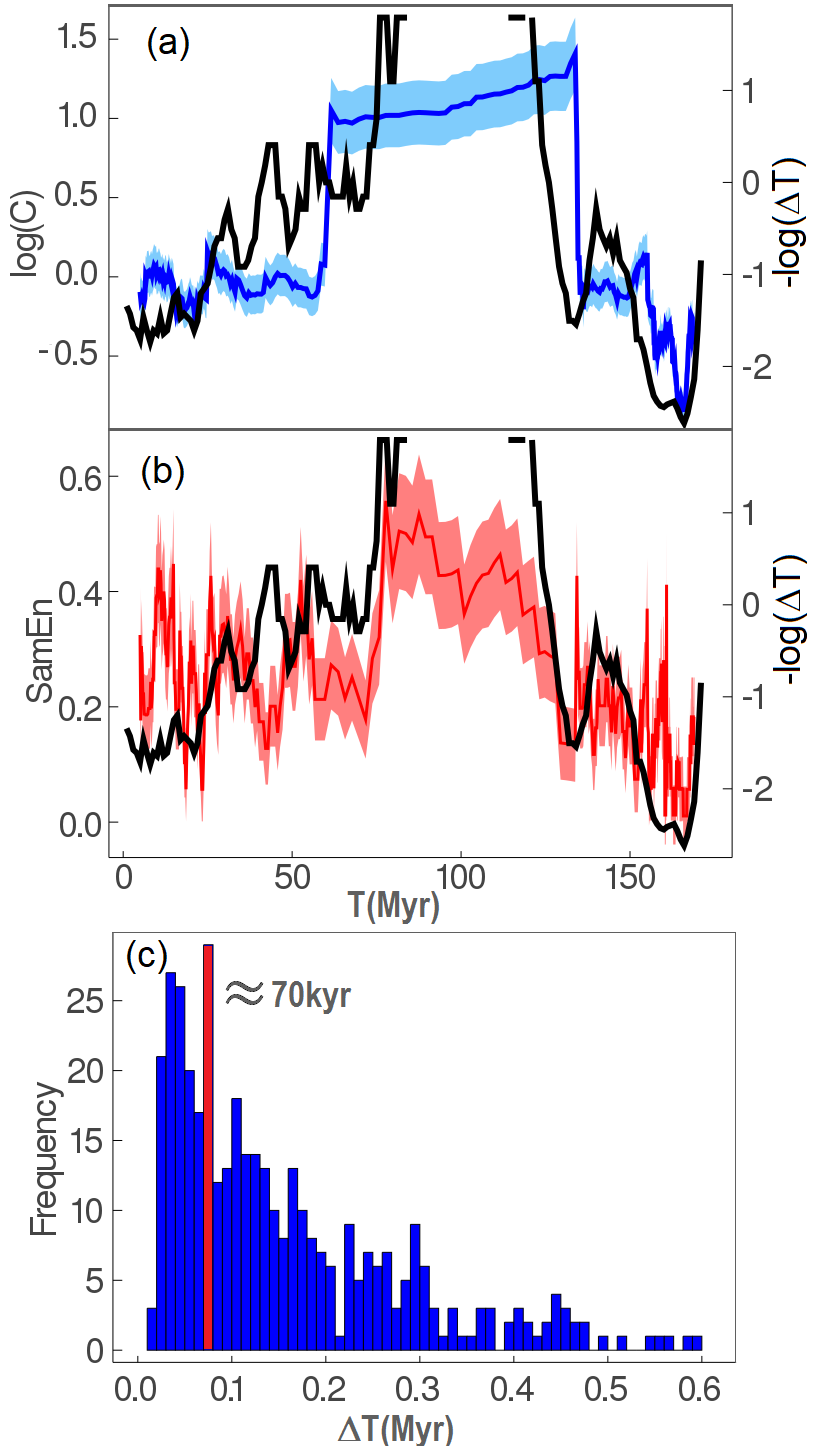}
\caption{(a) The blue curve represents coefficient of variation ($C$) calculated for the geomagnetic polarity scale. (b) The red curve stands for the sample entropy ($SamEn$) calculated for the geomagnetic polarity scale using parameters $r=0.2\sigma(X)$ and $m=2$. The black curve in (a) and (b) represents $-log(\Delta T)$ where $\Delta T$ corresponds to polarity intervals. Confidence intervals with 95\% confidence levels are calculated by bootstrap re-sampling. (c) Histogram of polarity intervals for $T\leq 600$ kyrs, highlighted in red the dominant $\approx 70$ kyrs intervals.}
\label{fig:data}
\end{figure}

In Figs. \ref{fig:data}(a) and \ref{fig:data}(b), we show $C$ and SamEn, respectively. These measures are evaluated from the geomagnetic reversals data for sequences of $30$ consecutive polarity intervals of duration $\Delta T$. These intervals are shown Figs. \ref{fig:data}(a) and \ref{fig:data}(b) as $-log(\Delta T)$ (black curve) for scaling purposes. There is an apparent inverse relationship between the regularity of the sequences of intervals and the corresponding reversal rate, with a particularly ordered behavior in the mid-Jurassic ($\approx$ $160$ Myrs ago). In Figure \ref{fig:data}(c), we show the distribution of intervals with sizes restricted to $\Delta T\leq 600$ kyrs. This histogram reveals a clustered character in the distribution of intervals with several peaks of occurrences, and the most frequent being $\Delta t \approx 70$ kyrs. A possible interpretation for these peaks is the occurrence of different dynamical regimes of the geodynamo. Such variability of regimes could be a result of changes in the core-mantle boundary \cite{zhang2011heat}, or even a stochastic resonance \cite{consolini2003stochastic} at which regularity is induced by the interaction of noise and external periodic forcing.

Moreover, the most frequent interval being $\Delta T \approx 70$ kyrs suggests that fast-reversing regimes are of high relevance for the geodynamo. This observation is compatible with the range of intervals found at around $155$\textendash$170$ Myrs ago, that are seemingly regular in accordance with Figs. \ref{fig:data}(a) and \ref{fig:data}(b). More recently, \citet{gallet2019extreme} reported a period with similar reversal rates at the Drumian (around $504.5$\textendash$500.5$ Myrs ago). Another interesting outcome of the analysis presented in Fig. \ref{fig:data}(a) is that $C$ can be viewed as a measure of the Poissonity of the distribution of reversal times, with $C \neq 1$ compatible with a large departure from Poissonity. This is the case in sequences in the vicinity of the Superchron non-reversing state ($\approx 80$\textendash$120$ Myrs) and in regular state near ($\approx 160$ Myrs). An important implication of this observation is the validity of models that describe the geomagnetic reversals as a stochastic exit problem with Poisson times \cite{buffett2018probabilistic,hoyng2004geomagnetic}, in this case $C$ could be used to asses the periods in the past when these models are compatible with the observations.

To illustrate how regularity could arise in the geodynamo, we use a reduced model for geomagnetic reversals introduced by Gissinger in Ref. \cite{Gissinger2010} and further analyzed in Ref. \cite{Gissinger2012}. It consists of a truncation of the magnetohydrodynamic equations that are consistent with the symmetry properties of the system. It retains only the dipole and quadruple components of the magnetic field, respectively $D$ and $Q$, coupled with a symmetry-breaking flow component $V$. Despite the simplicity of the model it presents an impressive resemblance with the qualitative behavior of the geodynamo and a surprisingly rich variety of regimes, including periodic windows, superchron-like non-reversing states, and irregular/chaotic reversal regimes. This model was used to perform data assimilation with coarse observations in order to estimate the expected time for the next dipole's reversal \cite{morzfeld2017coarse}. Compared to other simplifications, Gissinger's model had the best performance for this task. Moreover, estimates of the dimension of the reconstructed attractor associated to the geomagnetic axial dipole using embedding techniques on paleointensity data, \cite{Ryan2008}, indicate its dynamics is indeed low-dimensional ($3$-$7$ dimensions). The equation describing the model is given by:
\begin{eqnarray}
\nonumber
\dot{Q}&=&\mu Q-VD,\\  
\label{model}
\dot{D}&=&-\nu D+VQ,\\
\nonumber
\dot{V}&=&\Gamma - V+QD.
\end{eqnarray}
The parameters $\mu$, $\Gamma$ can be seen as forcing terms on the quadrupole $Q$ and velocity field $V$ components, while the parameter $\nu$ can be seen as an effective viscosity term leading to the dissipation of the dipole component $D$ of the geomagnetic field. By performing a bifurcation analysis of this model, Gissinger has already demonstrated the existence of intervals in the parameter $\mu$ for which the system exhibit stable periodic behavior, i.e. periodic windows (PWs) \cite{Gissinger2012}.

\begin{figure}[h]
\centering
\includegraphics[width=8cm,height=6.5cm]{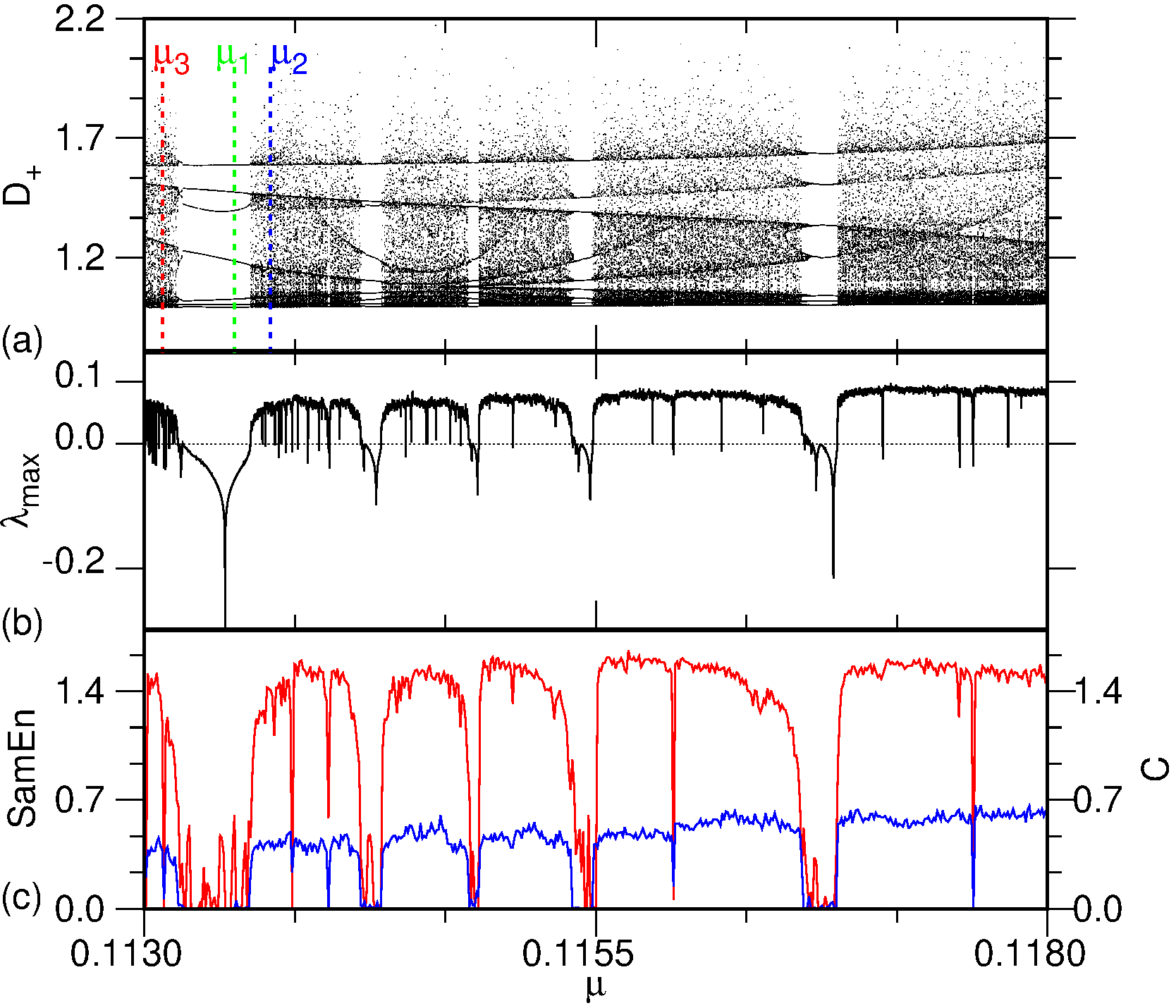}
\caption{(a) Bifurcation diagram of the Gissinger's model revealing parameter intervals leading to periodic behavior (PWs) and intervals leading to chaos. The bifurcation parameter $\mu$ is considered in the interval $[0.1130,0.1180]$. The green dashed line corresponds to $\mu_1=0.1135$, the blue dashed line indicates $\mu_2=0.1137$, and the red line $\mu_3=0.1131$. (b) The corresponding maximum Lyapunov exponents $\lambda_{max}$. (c) Blue represents coefficient of variation $C$ and red stands for the sample entropy $SamEn$.}.
\label{fig:Bif}
\end{figure}

Here, we associate the high levels of regularity observed in Fig. \ref{fig:data} to the occurrence of PWs in the model. Additionally, more than the stable periodic behavior within every PW, the chaotic behavior of parameter regions adjacent to them preserves vestiges of their periodicity. In order to illustrate these ideas, in Fig. \ref{fig:Bif}(a) we perform a bifurcation analysis of the system in a Poincar\'e section defined as $\Sigma=\{(Q,D,V) \in \mathcal{R}^3| aQ+bD=0 \}$ with $a=\sqrt{\mu+\Gamma\sqrt{\mu/\nu}}$ and $b=\sqrt{\nu+\Gamma\sqrt{\nu/\mu}}$. For $\mu \in [0.113,0.118]$, in Fig. \ref{fig:Bif}(a), we observe many PWs in a sequence, it is worth notice that even when the periodic orbits loose stability, becoming an unstable periodic orbit (UPO) and leading to chaotic behavior, the density of points continue to be higher near the dominant UPOs. In Fig. \ref{fig:Bif}(b), for same interval of $\mu$, we show the maximum Lyapunov exponents ($\lambda_{max}$), the PW corresponds to $\lambda_{max}<0$. Next, in Fig. \ref{fig:Bif}(c) we obtain the sample entropy (SamEn) and the variation coefficient (C) in the fashion of Fig. \ref{fig:data}. Both SamEn and C efficiently captures the occurrence of PWs in the model, this agreement supports the capability of these measures in detecting the regularities of reversals data reported in Fig. \ref{fig:data}.

In more detail, every PW in the bifurcation diagram shown in Fig. \ref{fig:Bif}(a) is delimited by a saddle-node bifurcation on its right side and a period-doubling route to chaos on its left side. In the chaotic regime occurring in the neighborhood of the saddle-node bifurcation (right side), a Pomeau-Manneville type-$I$ intermittency scenario \cite{eckmann1981roads, pomeau1980intermittent} gives rise to "ghost" limit cycles \cite{ruelle2014elements,medeiros2017trapping} with the same periodicity of the main orbit within the adjacent PW. On the PW's left side, the periodic orbits created in the PW continue to exist in an unstable way, as an UPO \cite{grebogi1988unstable}. Hence, for values of $\mu$ close enough to PWs on both sides, regularity will appear in the frequency of reversals. 

UPOs can be viewed as the skeleton of a chaotic attractor, so that statistical properties of the system can be interpreted in terms of the properties of the UPOs \cite{lucarini2019new,gritsun2017fluctuations}.
The influence of UPOs has been previously discussed in the context of reversal rates of the geomagnetic field in \citet{raphaldini2020evidence}. In this study, the authors show that the invariant probability distribution of the system can be well approximated by the dominant UPOs of the system. These orbits can be classified into two types: the global UPOs are the ones that perform a reversal (crossing the plane $D=0$), and the local ones, which are responsible for variations of the geomagnetic field without reversing in the polarity polarity. In particular during transitions to geomagnetic superchrons it is suggested that the set of global UPOs are destroyed.  Since UPOs are saddle orbits their stable manifold attracts the chaotic orbits that latter get expelled through the unstable one. For orbits sufficiently close to the stable manifold the motion mimics the UPO for extended times leaving a statistical imprint in the distribution function on the phase space. These almost periodic transients can latter be used to trace an initial approximation of the UPO that can be refined by iterative techniques involving the monodromy matrix.

\begin{figure}[h]
    \centering
    \includegraphics[width=1\linewidth]{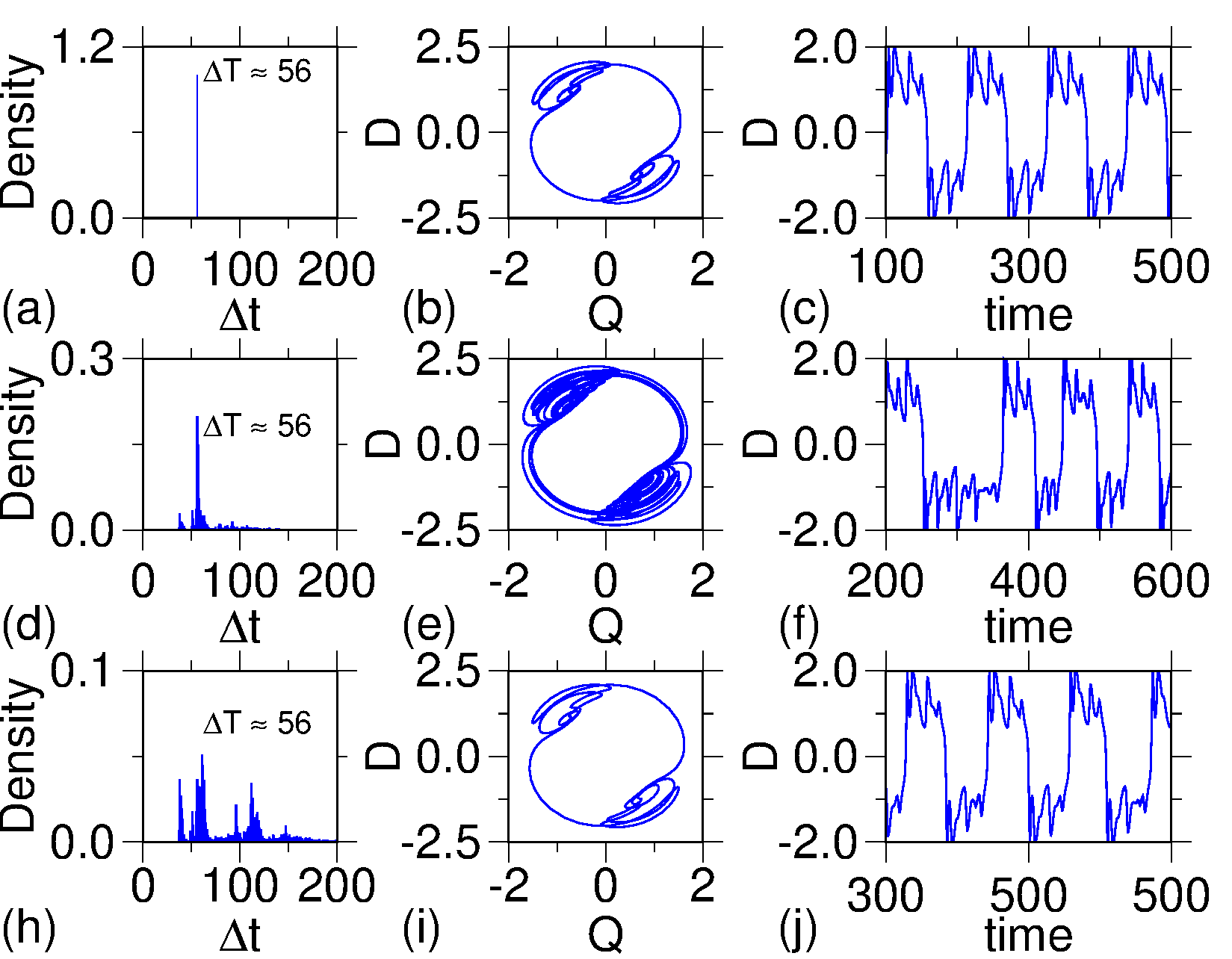}
    \caption{For different parameter settings, we show from left to right the distribution pf polarity intervals, a state-space ($Q \times D$) projection, and the time evolution of the model's dipole component (a)-(c) For $\mu_1=0.1135$, the system oscillates periodically in a limit cycle. (d)-(g) For $\mu_1=0.1137$ the model oscillates chaotically close to a "ghost" limit cycle. (h)-(j) For $\mu_1=0.1131$ the system is also chaotic, but an UPO dominates the dynamics.}
    \label{fig:model}
\end{figure}

To better clarify the occurrence of different degrees of regularity in the model, in Fig. \ref{fig:model}, we explore in detail its dynamical behavior for the three values of $\mu$ shown in Fig. \ref{fig:Bif}(a). For $\mu_1=0.1135$, the system oscillates in the most stable periodic orbit within a PW. In Fig. \ref{fig:model}(a), we show the respective distribution of time intervals $\Delta T$ for each polarity of the model's dipole component $D$. The sole peak indicates the periodicity of the stable limit cycle, i.e. $\Delta T \approx 56$. In Fig. \ref{fig:model}(b), we show a state-space projection ($ Q\times D$) of this limit cycle and, in \ref{fig:model}(c), we show the respective time evolution of the dipole component oscillating regularly. For $\mu_2=0.1137$, the system oscillates chaotically, yet, under the effect of the "ghost" limit cycle. In Fig. \ref{fig:model}(d), we show the distribution of polarity intervals $\Delta T$ for this parameter region. The high frequency of the polarity interval $\Delta T \approx 56$ indicates the high influence of the limit cycle nearby, i.e., its "ghost". The similarity between the stable limit cycle and its ghost can also be seen in some phases of the time evolution shown Fig. \ref{fig:model}(f). The state-space projection of such a chaotic attractor is shown in Fig. \ref{fig:model}(e). Finally, for $\mu_3=0.1131$, the reversal dynamics is also chaotic, however, it occurs in a parameter region where the stable periodic orbit of the neighbor PW exists in an unstable way, a UPO. The most frequent polarity interval in Fig. \ref{fig:model}(h) $\Delta T \approx 56$ indicates the dominance of such UPO corresponding to the PW's main periodic orbit but also presents peaks in multiples of the fundamental frequency which could be associated with the multi-peak structure presented in \ref{fig:data}(c). In Fig. \ref{fig:model}(i), we show an approximation of this dominant UPO. The time evolution of the dipole component $D$, shown in Fig. \ref{fig:model}(j), further illustrates the regular character of the oscillation close to the UPO.

In summary, based on measures of dispersion of sequences of consecutive chrons, we have demonstrated the existence of regularity in the geodynamo reversal process around $160$ million years ago. We have shown that at this epoch, these measures of dispersion, namely the coefficient of variation and the sample entropy, drop dramatically. The existence of periods with high reversal rates in the geodynamo has a very significant impact on the statistics of chron duration see histogram in Fig. \ref{fig:data}(c)). A fast reversing rate period of $\approx 70$ kyrs. corresponds to the largest peak in the histogram. As showed by \cite{carbone2006clustering}, there is a clustering pattern in the statistics of intervals that suggests the existence of different regimes in the geodynamo. This fast reversing and regular state are therefore highly relevant for the statistics of reversals.

The analogy with a simple model for the geomagnetic reversal, \cite{Gissinger2010}, shows that bifurcations of the system may lead to periodic windows in which the chaotic attractor collapses to a limit cycle. This is caused by either a period-doubling route or by a type-I Pomeu Manneville intermittency. We show that even in the chaotic regime when the parameters are close to a periodic window, the limit cycle ghost still impacts the dynamics leading to a large peak in the histogram of residence times, see Fig. \ref{fig:data}(c). This suggests that even when the geodynamo operates in a chaotic regime the effects of ghost limit cycles, or UPOs, can be felt, given that the system is close enough to a periodic window. Therefore, the large peak of $\approx$ $70$ kyrs. in  Fig. \ref{fig:data}(c) could either be the signature of a limit cycle itself, its "ghost", or its correspondent UPO.

The mechanism behind the variability of the duration of geomagnetic chrons is still elusive and it is, to a large extent, an open problem. Numerical geodynamo models suggest that the frequency of reversals depends crucially on the properties of the core-mantle boundary (CMB), in particular the heat flow in this interface \cite{amit2015lower}. In particular, the geometry of the distribution of heat flux in the core-mantle boundary may be of high importance for the geodynamo, with evidence showing correlations with the tectonic process \cite{petrelis2011plate}. An inverse relationship between the degree of dipolarity of the geomagnetic field and the reversal rates was found in \citet{franco2019paleomagnetic} which also coincides with the estimated heat flux at the CMB. Long term changes in this pattern may, therefore, alter the regime of operation of the dynamo.  

It's also important to determine whether the peak at the histogram with $\approx$ $70$ kyrs has some relation with external forcing or if it results from the nonlinear variability of the system solely. \citet{consolini2003stochastic} suggested that the $100$ kyrs. Millankovich orbital cycles could play a role in determining the frequency of reversals. In addition, they argued that stochastic resonance could also play a role and suggested that the distribution of polarity intervals is described as a superposition of Gaussians. Each one centred at multiples of a fundamental frenquency $T=100$ kyrs. This frequency has also been reported in the literature in inclination and intensity data \cite{yamazaki2002orbital}.

The hypothesis of the Earth's orbital dynamics as acting as one of the forcing mechanisms to the long-term geomagnetic field behavior remains a contentious subject in literature \citet{yamazaki2002orbital}. Some studies (e.g. \cite{Rochester1975}) advocated that the precessionally-driven energy supply would not be significant to empower the outer core/mantle relative motion. However, it has been questioned more recently by other authors (e.g., \citet{christensen2004power}), indicating that the effect of orbital forcing on the geodynamo's energy budget cannot be ruled out. In the hypothesis of an orbitally-driven origin of the $70$ kyrs signal, it cannot be straightforwardly attributed to one of the Earth's orbital parameters. Nevertheless, a similar multimillennial-scale quasi-periodicity has been reported as possibly resulted from distinctive processes – for instance, as a transient frequency associated with the eccentricity-precession modulation \cite{hinnov2000new}, as well as a short-eccentricity cycle expression during the minimal amplitude of the long-eccentricity cycle \cite{hennebert2012hunting}. Whether the $T\approx 70$ kyrs. arises purely from the internal geodynamo dynamics or some self-organizing/resonance effect with external orbital forcing, as in \citet{consolini2003stochastic}, remains to be investigated in future observational, theoretical and numerical studies.

This work was supported by FAPESP (Processes: 	2017/23417-5, 2018/03211-6, 2013/26598-0). ESM acknowledges support by the Deutsche Forschungsgemeinschaft (DFG, German Research Foundation) - Projektnummer - 163436311 - SFB 910.


%

\end{document}